\documentclass[twocolumn]{aastex631}

\shorttitle{\Lya Halos around [O III]-Selected Galaxies in HETDEX}
\shortauthors{Lujan Niemeyer et al.}

\graphicspath{{./}{figures/}}

\usepackage{amsmath}
\usepackage{xspace}
\newcommand{\angstrom}{\textup{\AA}}
\newcommand{\Lya}{Ly$\alpha$\xspace}
\newcommand{\oiii}{[\ion{O}{3}]\xspace}
\newcommand{\HI}{\ion{H}{1}\xspace}

\newcommand{\numgal}{1034\xspace}
\newcommand{\sblya}{\Lya SB\xspace}

\newcommand{\threedhst}{3D-HST\xspace}

\newcommand{\hetdex}{HETDEX}
\newcommand{\sbunits}{\mathrm{erg}\,\mathrm{s}^{-1}\,\mathrm{cm}^{-2}\,\mathrm{arcsec}^{-2}}

\defcitealias{lujanniemeyer/etal2022}{LN22}

\begin{document}

\title{\Lya Halos around [O III]-Selected Galaxies in HETDEX}

\author[0000-0002-6907-8370]{Maja Lujan Niemeyer}
\affiliation{Max-Planck-Institut f\"{u}r Astrophysik, Karl-Schwarzschild-Str. 1, 85741 Garching, Germany}
\email{maja@mpa-garching.mpg.de}

\author[0000-0003-4381-5245]{William P. Bowman}
\affiliation{Department of Astronomy \& Astrophysics, The Pennsylvania State University, University Park, PA 16802, USA}
\affiliation{Institute for Gravitation and the Cosmos, The Pennsylvania State University, University Park, PA 16802, USA}

\author[0000-0002-1328-0211]{Robin Ciardullo}
\affiliation{Department of Astronomy \& Astrophysics, The Pennsylvania State University, University Park, PA 16802, USA}
\affiliation{Institute for Gravitation and the Cosmos, The Pennsylvania State University, University Park, PA 16802, USA}

\author[0000-0003-2491-060X]{Max Gronke}
\affiliation{Max-Planck-Institut f\"{u}r Astrophysik, Karl-Schwarzschild-Str. 1, 85741 Garching, Germany}

\author[0000-0002-0136-2404]{Eiichiro Komatsu}
\affiliation{Max-Planck-Institut f\"{u}r Astrophysik, Karl-Schwarzschild-Str. 1, 85741 Garching, Germany}
\affiliation{Kavli Institute for the Physics and Mathematics of the Universe (Kavli IPMU, WPI), University of Tokyo, Chiba 277-8582, Japan}

\author[0000-0002-7025-6058]{Maximilian Fabricius}
\affiliation{Max-Planck-Insitut f\"ur Extraterrestrische Physik, Gie{\ss}enbachstra{\ss}e 1, 85748 Garching, Germany}
\affiliation{Universit\"ats-Sternwarte M\"unchen, Scheinerstra{\ss}e 1, D-81679 München, Germany}

\author[0000-0003-2575-0652]{Daniel J. Farrow}
\affiliation{Max-Planck-Insitut f\"ur Extraterrestrische Physik, Gie{\ss}enbachstra{\ss}e 1, 85748 Garching, Germany}
\affiliation{Universit\"ats-Sternwarte M\"unchen, Scheinerstra{\ss}e 1, D-81679 München, Germany}

\author[0000-0001-8519-1130]{Steven L. Finkelstein}
\affiliation{Department of Astronomy, The University of Texas at Austin, 2515 Speedway Boulevard, Austin, TX 78712, USA}

\author[0000-0002-8433-8185]{Karl Gebhardt}
\affiliation{Department of Astronomy, The University of Texas at Austin, 2515 Speedway Boulevard, Austin, TX 78712, USA}

\author[0000-0001-6842-2371]{Caryl Gronwall}
\affiliation{Department of Astronomy \& Astrophysics, The Pennsylvania State University, University Park, PA 16802, USA}
\affiliation{Institute for Gravitation and the Cosmos, The Pennsylvania State University, University Park, PA 16802, USA}

\author[0000-0001-6717-7685]{Gary J. Hill}
\affiliation{Department of Astronomy, The University of Texas at Austin, 2515 Speedway Boulevard, Austin, TX 78712, USA}
\affiliation{McDonald Observatory, The University of Texas at Austin, 2515 Speedway Boulevard, Austin, TX 78712, USA}

\author[0000-0001-5561-2010]{Chenxu Liu}
\affiliation{Department of Astronomy, The University of Texas at Austin, 2515 Speedway Boulevard, Austin, TX 78712, USA}

\author[0000-0002-2307-0146]{Erin Mentuch Cooper}
\affiliation{Department of Astronomy, The University of Texas at Austin, 2515 Speedway Boulevard, Austin, TX 78712, USA}
\affiliation{McDonald Observatory, The University of Texas at Austin, 2515 Speedway Boulevard, Austin, TX 78712, USA}

\author[0000-0001-7240-7449]{Donald P. Schneider}
\affiliation{Department of Astronomy \& Astrophysics, The Pennsylvania State University, University Park, PA 16802, USA}
\affiliation{Institute for Gravitation and the Cosmos, The Pennsylvania State University, University Park, PA 16802, USA}

\author[0000-0002-7327-565X]{Sarah Tuttle}
\affiliation{Department of Astronomy, University of Washington, Seattle}

\author[0000-0003-2307-0629]{Gregory R. Zeimann}
\affiliation{Hobby-Eberly Telescope, University of Texas, Austin, Austin, TX, 78712, USA}


\begin{abstract}
We present extended Lyman-$\alpha$ (Ly$\alpha$) emission out to $800$~kpc of \numgal \oiii-selected galaxies at redshifts $1.9<z<2.35$ using the Hobby-Eberly Telescope Dark Energy Experiment (HETDEX\null). The locations and redshifts of the galaxies are taken from the 3D-HST survey. The median-stacked surface brightness profile of \Lya emission of the \oiii-selected galaxies agrees well with that of 968 bright \Lya-emitting galaxies (LAEs) at $r>40$~kpc from the galaxy centers. The surface brightness in the inner parts ($r<10$~kpc) around the \oiii-selected galaxies, however, is ten times fainter than that of the LAEs. 
Our results are consistent with the notion that photons dominating the outer regions of the \Lya halos are not produced in the central galaxies but originate outside of them.
\end{abstract}


\section{Introduction} \label{sec:intro}

Extended Lyman-$\alpha$ (Ly$\alpha$) emission around star-forming galaxies without an active galactic nucleus (AGN) has been found around Lyman-break galaxies \citep[LBGs; e.g.;][for a review on LAEs]{steidel/etal:2011,kusakabe/etal:2022,ouchi:2019} and Ly$\alpha$-emitting galaxies \citep[LAEs; e.g.;][]{wisotzki/etal:2016,kikuchihara/etal:2021}. 
One source of \Lya photons is the local recombination of hydrogen atoms ionized by photons from young, massive stars in star-forming regions. After their escape from the interstellar medium (ISM), \Lya photons will be scattered by neutral hydrogen atoms in the circumgalactic medium (CGM) and intergalactic medium (IGM\null). Hydrogen atoms in the CGM and IGM can also be ionized by photons from more distant AGN or star-forming regions, called the ultraviolet (UV) background, and recombine to emit \Lya photons \citep[``fluorescence''; e.g.;][]{gould/weinberg:1996}. \Lya photons from satellite galaxies \citep{mas-ribas/etal:2017} and collisional excitation of hydrogen atoms in cooling gas \citep[``cooling radiation''; e.g.;][]{haiman/spaans/quataert:2000} can add to the extended \Lya emission. 

Because the contribution of scattered photons from the central galaxy to the halo depends on the galaxy's \Lya emission, comparing the \Lya surface brightness (SB) profiles of galaxies with different intrinsic \Lya luminosities or escape fractions can probe the origin of \Lya halos.
Because LAEs are selected using their large \Lya equivalent width (EW), they comprise a biased subset of high-redshift galaxies that have a large \Lya escape fraction along the line of sight (LOS).
Galaxies selected via other methods such as LBGs or via their rest-frame optical emission lines may have similar physical properties, but with smaller \Lya escape fractions than LAEs.
\citet{erb2016}, \citet{hathi/etal:2016}, \citet{trainor/etal:2016,trainor/etal:2019}, and \citet{reddy/etal:2022} argue that LAEs have different properties than other star-forming galaxies, such as less dust and metal content, lower star-formation rates (SFR) and stellar masses, and higher \HI covering fractions.
Conversely, \citet{hagen/etal:2016} and \citet{shimakawa2017} report no statistical difference between the properties of the samples of LAEs and rest-frame-optical emission-line galaxies except at high stellar masses.
Hence, comparing the \Lya halo profiles of LAEs with those of rest-frame-optical emission line galaxies can shed light on the emission sources and mechanisms of \Lya halos.

We compare the median-stacked \sblya profile of \numgal galaxies at $1.9<z<2.35$ selected via their rest-frame optical emission lines in the \threedhst\ survey \citep{brammer2012, momcheva2016, bowman2019,bowman2020} with that of LAEs at $1.9<z<3.5$ detected in the Hobby-Eberly Telescope Dark Energy Experiment \citep[HETDEX;][hereafter LN22]{hill/etal:2021,gebhardt/etal:2021,lujanniemeyer/etal2022}. 
We use integral-field spectroscopic data from HETDEX to extract the \sblya profiles.

We adopt a flat $\Lambda$-cold-dark-matter cosmology with $H_0=67.37\,\mathrm{km}\,\mathrm{s}^{-1}\,\mathrm{Mpc}^{-1}$ and $\Omega_{\mathrm{m},0}=0.3147$ \citep{planck/etal:2018}. All distances are in units of physical kpc unless noted otherwise.

\section{Data and Galaxy Samples}\label{sec:data}

\subsection{HETDEX Data}
We use spectra from the HETDEX survey \citep[][]{gebhardt/etal:2021}, specifically the internal data release 3.
The survey uses the VIRUS instrument on the $10-$m Hobby-Eberly Telescope (HET). See \citet{hill/etal:2021} for details.

VIRUS consists of up to $78$ integral-field unit fiber arrays (IFUs), each of which contains $448$ $1.5''$-diameter fibers and covers $51''\times 51''$ on the sky. The fibers from each IFU are fed to a low-resolution ($R\simeq 800$) spectrograph covering $3500-5500\,\angstrom$.
The IFUs with $\simeq 35$k total fibers are distributed on a grid with $100''$ spacing throughout the $18'$ diameter of the telescope's field of view. Each HETDEX observation comprises three 6-minute exposures, which are dithered to fill in gaps between the fibers. Because the gaps between the IFUs remain in an individual observation, the filling factor is $\simeq 1/4.6$. 

We use the full-frame sky-subtracted data (details in \citet{gebhardt/etal:2021}; \citetalias{lujanniemeyer/etal2022}).
This sky-subtraction method measures the sky emission from the entire $18'$-diameter field of view of VIRUS to ensure that extended emission on the scale of an IFU or larger is not removed along with the sky model.
The full-frame sky subtraction in the internal HETDEX DR 3 has some differences to that in DR 2, which is used in \citetalias{lujanniemeyer/etal2022}.
Instead of roughly $75\%$ of the total fibers with the lowest continuum emission, only $50\%$ are used for the sky estimate. This helps prevent the oversubtraction of continuum emission due to unresolved sources. To be more conservative, the smooth background subtraction within a six fibers by $600\,\angstrom$ window is omitted.
These changes do not affect our measurement because we perform a local continuum subtraction. As expected, the \sblya profiles of the LAEs using the data from DR 3 and DR 2 and the same stacking procedure are very similar.
We mask the wavelength regions around the brightest sky emission lines to avoid residuals associated with this component.  

\subsection{\oiii-Galaxy Sample}
Our \oiii-galaxy sample is drawn from \threedhst\ \citep{brammer2012, momcheva2016}, 
an \textsl{HST} Treasury program which used 2-orbit exposures with the WFC3 G141 grism to 
observe \mbox{$\simeq 625$ arcmin$^2$} of sky within the Cosmic Assembly
Near-IR Deep Extragalactic Legacy Survey \citep[CANDELS;][]{grogin2011,
koekemoer2011} footprint. \citet{bowman2019} vetted this dataset
to define a sample of $\simeq 2000$ optical-emission line galaxies with
IR continuum magnitude $m_{\rm J+JH+H} < 26$, unambiguous emission-line
redshifts between $1.90<z<2.35$, and a 50\% line-flux completeness limit of
$\sim 4 \times 10^{-17}$~ergs~cm$^{-2}$~s$^{-1}$. In over 90\% of the sample,
the brightest emission line in the spectral region surveyed by the grating is \oiii
$\lambda 5007$; in 90\% of the remaining galaxies, [\ion{O}{2}] dominates.
Most AGN have been removed from this dataset via comparisons
with X-ray source catalogs, and \citet{bowman2019} estimate the fraction of
remaining AGN to be less than 5\%.

More than half of the \citet{bowman2019} sample has been surveyed as part of
science verification for the HETDEX survey \citep{gebhardt/etal:2021}; this dataset
includes over 900 galaxies that have been observed more than once, with some being observed up to 15 times. These repeat observations partly cover the gaps between IFUs and provide a better spatial sampling of the datacube.
\citet{weiss2021} measured the mean \Lya escape fraction of the subsample of these galaxies
present in HETDEX DR2 ($6^{+0.6}_{-0.5}\%$) and determined the systematic behavior of the \Lya escape
versus stellar mass, SFR, internal extinction, half-light
radius, and excitation.

We only include HETDEX observations with good seeing (point-spread function (PSF) full-width-at-half-maximum $<1\farcs 7$) and observing conditions (total system throughput $>0.1=$ 13th percentile). These requirements are less strict than for the LAE sample because too few observations of \oiii galaxies meet these requirements. We inspect the remaining observations and exclude data with obvious artifacts such as interference patterns.
We require that for an \oiii galaxy's halo to be included in our analysis, the center of the galaxy must lie within $3''$ of the center of a HETDEX fiber. A total of \numgal \oiii galaxies (in 44 HETDEX observations) meet our selection criteria, with
$57$ ($\simeq 6$\%) having a \Lya detection in HETDEX (within $3''$ and $15\,\angstrom$ of the expected emission line).
Each galaxy was observed in $1-15$ separate observations; thus there are $7401$ individual observations of the \numgal galaxies. Their mean redshift is $\langle z\rangle = 2.1$.

Because of the abundance of imaging data in the CANDELS fields, the physical properties of our \oiii 
sample have been well characterized, with stellar masses between $8.2 \lesssim \log_{10}
M/M_{\odot} \lesssim 11.4$ (median mass of $\log_{10} M/M_{\odot} = 9.3$),
SFRs between $0.02 \lesssim \textrm{SFR} \lesssim 250 \, M_{\odot}$~yr$^{-1}$
(median value of $1.9 M_{\odot}$~yr$^{-1}$),
internal extinctions between $0 \lesssim E(B-V) \lesssim 0.6$ (median
of $E(B-V) = 0.09$), and optical half-light radii
$R_e \lesssim 5$~kpc (with a median of $1.4$~kpc).
The full distribution of these
properties, along with their \oiii luminosity function and equivalent width
distribution, can be found in \citet{bowman2019,bowman2020,bowman/etal:2021}.

To study the potential dependence of the \sblya profile on various galaxy properties, we form two sub-samples above and below the median observed $L_{[\rm O\,III]}$ ($41.3\lesssim \log_{10} L\,\mathrm{erg^{-1}\,s}\lesssim 43.1$, median $42.1$, 517/517 sources above/below), SFR (517/517 sources above/below), stellar mass (517/517 sources above/below), H$\beta$ flux ($\lesssim 8\times10^{-17}\,\mathrm{erg\,s^{-1}\,cm^{-2}}$, median $10^{-17}\,\mathrm{erg\,s^{-1}\,cm^{-2}}$, 516/518 sources above/below), [O II] flux ($\lesssim 1.7\times10^{-16}\,\mathrm{erg\,s^{-1}\,cm^{-2}}$, median $2\times10^{-17}\,\mathrm{erg\,s^{-1}\,cm^{-2}}$, 517/517 sources above/below), dust attenuation (515/519 sources above/below), and UV luminosity ($25.4\lesssim \log_{10} L_{1600}\,\mathrm{erg^{-1}\,s\,Hz}\lesssim 29.4$, median $28.5$, 517/517 sources above/below).
We also fit a line to the SFR as a function of stellar mass and create subsamples above and below this linear relation (459/575 sources above/below). We omitted unrealistic values from the spectral-energy-distribution fits in the property ranges above.

We estimate the virial radius of the host dark matter halos using the stellar mass-halo mass relation of \citet{behroozi/etal:2019}. Roughly $68\%$ of the galaxies in our \oiii sample have stellar masses between $10^{8.8}\,M_\odot$ and $10^{9.9}\,M_\odot$ and therefore reside in $10^{11.4}\,M_\odot$ to $10^{11.9}\,M_\odot$ dark matter halos. Following the definition of $r_{\rm vir}$ of \citet{bryan/norman:1998}, we obtain $r_{\rm vir}\simeq59-105\,\mathrm{kpc}$.

\subsection{LAE Sample}

The LAE sample is selected from the \hetdex\ survey and is described in \citetalias{lujanniemeyer/etal2022}.
It consists of $968$ LAEs at $1.9<z<3.5$ with narrow lines (\Lya line $\mathrm{FWHM}<1000\,\mathrm{km}\,\mathrm{s}^{-1}$) and \Lya luminosities $10^{42.4}\,\mathrm{erg}\,\mathrm{s}^{-1}\lesssim L_{\rm Ly\alpha}< 10^{43}\,\mathrm{erg}\,\mathrm{s}^{-1}$. These conditions remove most AGN from the sample. Each LAE was observed once. $364$ of these LAEs are at $z<2.35$. The equivalent widths of the Ly$\alpha$ and other lines measured from the median-stacked rest-frame spectrum are consistent with star formation being the main powering mechanism of the \Lya emission.
To resolve \Lya halos, \citetalias{lujanniemeyer/etal2022} chose LAEs observed with PSF $\mathrm{FWHM}<1.4''$ throughput $>0.13=$ 40th percentile.
While we do not know the SFR and stellar mass of this sample, LAEs at $z\simeq 2.2$ with slightly lower \Lya luminosity ($L_{\rm Ly\alpha}\simeq 10^{42.3}\rm\, erg\,s^{-1}$) typically have ${\rm SFR}\simeq 14\,M_\odot\,\rm yr^{-1}$ and stellar mass $M_\star\simeq 5\times 10^8\,M_\odot$ \citep[][]{nakajima/etal:2012}. HETDEX LAEs have similar SFR and stellar mass (\nolinebreak{McCarron} et al. 2022, ApJ submitted).

\section{Measurement of \Lya Halos}\label{sec:methods}

\subsection{Extraction of Surface Brightness and Stacking}\label{subsec:sb_extraction_and_stacking}

We measure the \Lya halo profiles of the \oiii galaxies and around our comparative sample of LAEs following a similar procedure to \citetalias{lujanniemeyer/etal2022}. The LAE profiles are consistent with each other.
Because we do not detect individual \Lya emission lines of most \oiii galaxies, we assume that the observed \Lya line lies at the \threedhst redshift.
First, we remove continuum emission from the spectra to isolate \Lya from the continuum flux and to mitigate the impact of continuum emission from projected neighbors. From each fiber spectrum, we subtract the median flux between 11.7 and 40\,\AA\ (observed) away from the \Lya line on the red and blue side.
We then integrate the flux around the expected \Lya wavelength, obtaining a surface brightness for each fiber.
We choose an integration window of $\lambda^{\rm obs}_{\rm Ly\alpha}\pm 10\,\angstrom$ to account for the uncertainty of the expected observed \Lya wavelength due the limited spectral resolution of the grism ($R\simeq130$).

The \Lya line can be redshifted by $\simeq 200\,\mathrm{km}\,\mathrm{s}^{-1}$ from a galaxy's redshift because of radiative transfer effects \citep[e.g.,][]{nakajima/etal:2018b}. We tested redshifting the integration window by $200\,\mathrm{km}\,\mathrm{s}^{-1}$ and subtracting the continuum on the red side of the shifted \Lya line. We also tested subtracting only the red continuum without shifting the line. Both tests produce a \sblya profile consistent with our results. 

We define two LAE samples, the entire sample and the low-redshift ($z<2.35$) subsample. For the comparison with the entire LAE sample we correct for cosmological surface brightness dimming to the mean redshift (2.1) of the \oiii galaxy sample (factor $(1+z)^4\times(1+2.1)^{-4}$).
For the comparison with previous results (\S~ \ref{subsec:comparison_with_previous_results}) we convert the surface brightness of each LAE and \oiii galaxy to surface luminosity to account for cosmological surface brightness dimming. The surface luminosity ${\rm SL}_{\rm Ly\alpha}$ relates to the surface brightness ${\rm SB}_{\rm Ly\alpha}$ as ${\rm SL}_{\rm Ly\alpha} = dL / dA_{\rm em} = 4\pi \, (1+z)^4\, {\rm SB}_{\rm Ly\alpha}$, where $dA_{\rm em}$ is the surface area at emission.

We sort the fibers around each galaxy in each observation by their distance from the galaxy center, and place them in radial bins with the bin edges at $5, 10, 15, 20, 30, 40,60,80,160,320,$ and $800\,\mathrm{kpc}$.

We take the median of all fibers within each bin around each galaxy observation individually. Then we take the median of these radial profiles and estimate the uncertainty via a bootstrap algorithm.
At least $355$ ($527$) \oiii galaxies (galaxy observations) contribute to each bin.

\subsection{Estimating Systematic Uncertainty}
We estimate the systematic uncertainty in two parts following the approach of \citetalias{lujanniemeyer/etal2022}.
The first estimates the background surface brightness at the same wavelengths and in the same observations as the galaxies separately for each galaxy sample. This background includes physical emission, e.g. from interlopers, sky emission residuals, and systematics introduced in the continuum subtraction. We calculate the median surface brightness of all fibers farther away than $800\,\mathrm{kpc}$ from each galaxy observation. The median of these determines the background and a bootstrap algorithm determines the uncertainty. We find $(1.51\pm 0.02)\times 10^{-19}\sbunits$ for the \oiii galaxies. We subtract this background from the median profile of the galaxies.

The second part determines the systematic uncertainty of the median radial profile in proximity of the galaxies. We repeat the stacking procedure $40$ times, but with the central wavelength shifted between $20\,\angstrom$ and $210\,\angstrom$ in increments of $10\,\angstrom$ in both wavelength directions. Some galaxies at the blue end of the covered wavelength range have fewer than $40$ wavelength-shifted profiles.
The standard deviation per bin of these $40$ Ly$\alpha$-free profiles determines the systematic uncertainty of the median \sblya profile. The median ratio of this systematic uncertainty to the uncertainty from the bootstrap algorithm is $1.4$. In each bin, we choose the larger of the two estimates as the final uncertainty.
The mean and median of the wavelength-shifted profiles are consistent with the background surface brightness.

We stack the radial profiles of stars from the Gaia DR 2 \citep[][]{gaiadr2/2018} 
in the same manner as \citetalias{lujanniemeyer/etal2022} out to $100''$. The median profile plateaus at $10''<r<100''$, presumably because of unmasked continuum sources and the lack of a continuum and background subtraction. We subtract the mean value at $r>10''$. We obtain a separate star profile for the observations of LAEs and \oiii emitters and scale them to match the flux within $2''$ of the galaxy profiles.
Both profiles are modeled well by a Moffat function with $\beta=2.2$ with the mean seeing FWHM as the observations.

\section{Results and Discussion}\label{sec:results}
\begin{figure}
    \centering
    \includegraphics[width=0.47\textwidth]{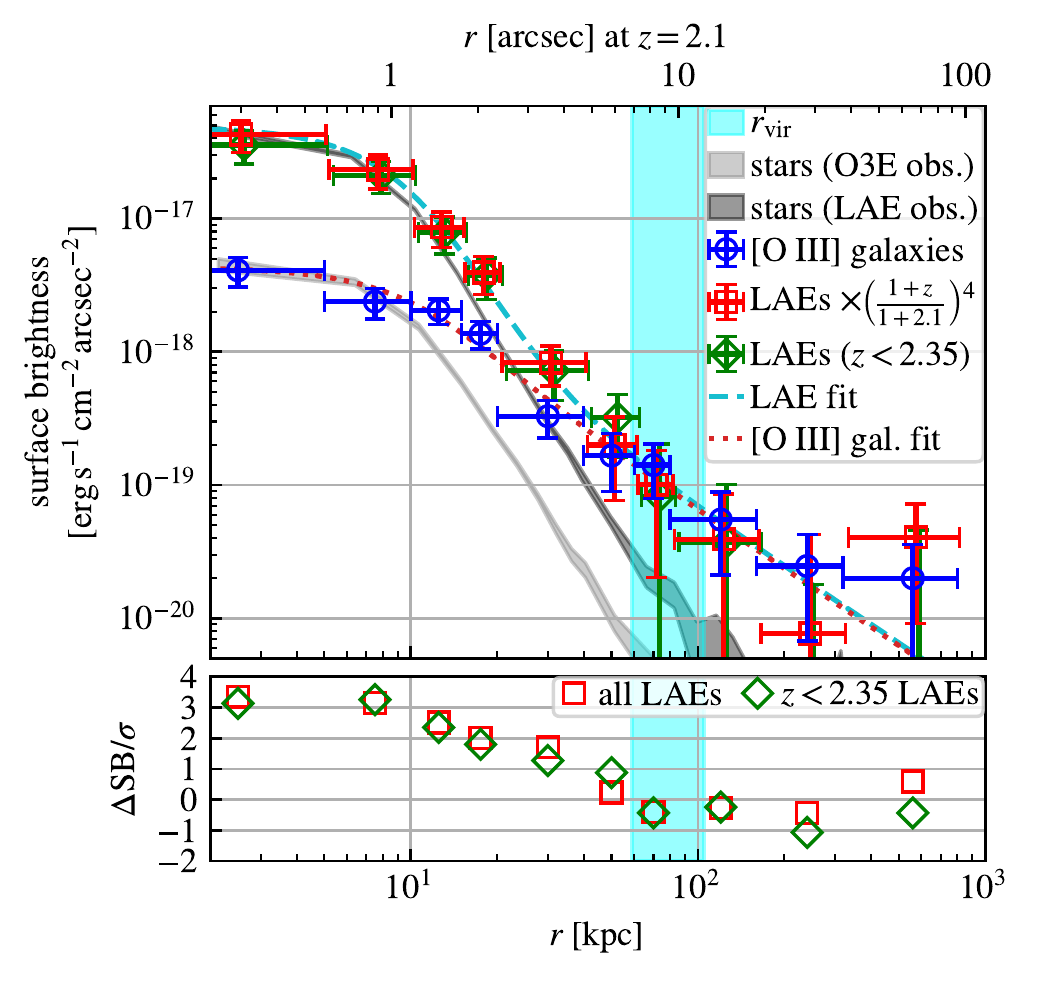}
    \caption{\textbf{Top}: Median \sblya profile of \numgal \oiii galaxies (blue circles) compared to the redshift-adjusted profile of all LAEs (red squares) and the profile of the LAEs at $z<2.35$ (green diamonds). The LAE profiles are slightly shifted along the x axis for better visibility. The star profile in the \oiii galaxy observations at $z=2.1$ and that in the LAE observations at $z=2.5$ are shown as light gray and dark gray areas, respectively. The cyan area shows the estimated virial radius of the host dark matter halos of the \oiii galaxies. The dotted and dashed lines show the best-fit PSF-plus-powerlaw model.
    \textbf{Bottom}: Significance of the difference between the profiles, i.e. the difference between the \oiii profile and the LAE profiles divided by the uncertainties added in quadrature, with the same symbols as above.}
    \label{fig:oiii_lya_profile_vs_laes}
\end{figure}

\begin{figure*}
 \centering
    \includegraphics[width=0.95\textwidth]{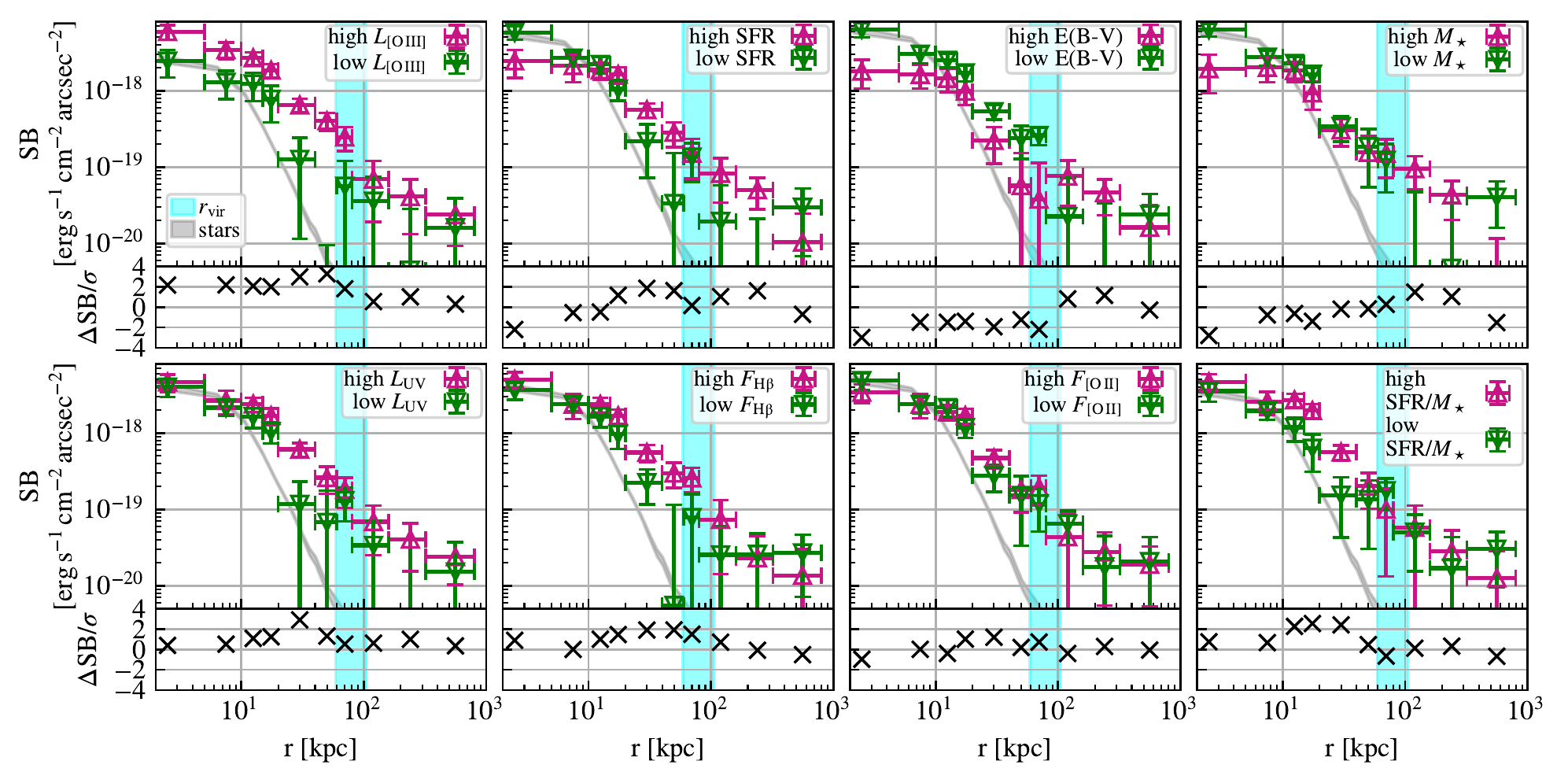}
    \caption{Median \sblya profiles of differently separated subsamples. Except for the bottom right, each panel includes the subsample above (upward-facing magenta triangles) and below (downward-facing green triangles) the median of one property. 
    The bottom right panel shows the subsample above and below the linear SFR-$M_\star$ relation.
    The bottom part of each panel shows the significance of the difference between the two profiles.
    }
    \label{fig:many_cuts}
\end{figure*}
\begin{figure}
    \centering
    \includegraphics[width=0.47\textwidth]{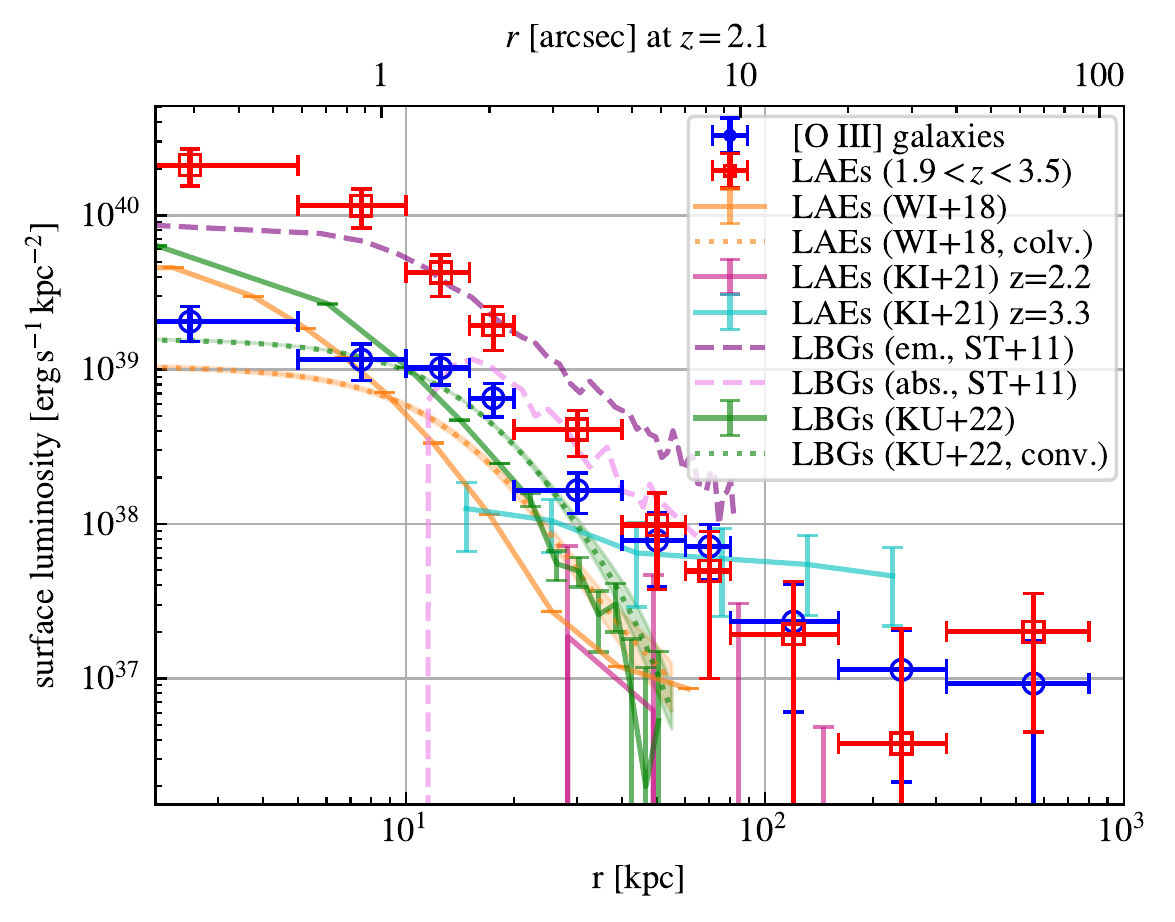}
    \caption{Comparison of the surface luminosity profile of the \oiii galaxies (blue circles) and that of the LAEs (red squares) with LAEs at $z=3-4$ \citep[WI+18;][]{wisotzki/etal:2018}, LAEs at $z=2.2$ and $z=3.3$ \citep[KI+21;][]{kikuchihara/etal:2021}, LBGs with net Lya emission and absorption at $z=2.65$ \citep[ST+11;][]{steidel/etal:2011}, and LBGs at $z=3.56$ \citep[KU+22;][]{kusakabe/etal:2022}. The dotted profiles are convolved with the VIRUS PSF.}
    \label{fig:literature_comparison}
\end{figure}

Figure \ref{fig:oiii_lya_profile_vs_laes} presents the median \sblya profile of the \oiii galaxies out to $800\,\mathrm{kpc}$.
The profile is significantly more extended than the star profile.
Figure \ref{fig:oiii_lya_profile_vs_laes} also shows the median redshift-adjusted \sblya profile of the LAE sample at $1.9<z<3.5$. While the \oiii-galaxy profile is an order of magnitude fainter at $r<10\,\mathrm{kpc}$, it reaches a consistent surface brightness at $r>40\,\mathrm{kpc}$. The profiles of the entire LAE sample and the subsample at $z<2.35$ are consistent at all radii.

\citet{byrohl/etal:2021} find in their simulation that the photons in the core predominantly originate from the central galaxy, but those at large distances originate from other galaxies. Hence the central surface brightness should depend on the amount of photons escaping the central galaxy. At large distances, however, galaxies with similar CGM and clustering properties should have similar \sblya profiles.

While the intrinsic \Lya luminosities of the galaxy samples are unknown, the small \Lya escape fraction of the \oiii galaxies along the LOS can explain the lower surface brightness of the profile in the core. In contrast, \citet{leclercq/etal:2017} find a weak positive correlation between the halo scale length and \Lya luminosity of the inner halo, implying that brighter LAEs have flatter halos.
The similarity of the \sblya profiles of the two galaxy samples at $r>40\,\mathrm{kpc}$ supports the picture in which the outer parts of the profiles are dominated by photons not related to \Lya emission produced in the central galaxy.
The profiles are modeled well by a PSF-plus-powerlaw model, with power law index $-1.45$, cut off at $r_{\rm min}\simeq 2\,\mathrm{kpc}$.
While the power-law component is designed to be identical, the best-fit PSF component of the LAE profile is $40$ times brighter than that of the \oiii galaxies.

Figure \ref{fig:many_cuts} shows the median \sblya profiles of several subsamples of the \oiii galaxies. Most profiles are similar ($<2\sigma$ difference). The following differences are statistically significant ($>2\sigma$).
The \Lya surface brightness of the low-$L_{[\rm O\,III]}$ sample is lower than that of the high-$L_{[\rm O\,III]}$ sample at $r<60\,\rm kpc$. While the low-$L_{[\rm O\,III]}$ profile follows the star profile out to $40\,\mathrm{kpc}$, it increases to match the high-$L_{[\rm O\,III]}$ profile at $r>60\,\mathrm{kpc}$.
The subsamples with high dust attenuation, stellar mass, and SFR are fainter at $r<5\,\rm kpc$ than those with low dust attenuation, stellar mass, and SFR, but similar at larger distances. This can be explained by lower escape fractions for those subsamples, consistent with the notion that the escape fraction anti-correlates with dust extinction, stellar mass, and SFR \citep[][]{runnholm/etal:2020,weiss2021}.
The profiles of the subsamples with low UV luminosity or below the SFR-stellar mass relation are similar to those with high UV luminosity or above the SFR-stellar mass relation at most radii, but fainter at intermediate distances, similar to the low-$L_{\rm [O III]}$ subsample.
This suggests that the surface brightness is independent of the properties of the central galaxies at large distances.
However, the uncertainties at large radii are large and more data are necessary for a clear conclusion.
The similarity of the \sblya profiles at different stellar masses appears to contradict the result of \citet[][]{byrohl/etal:2021}. Their fiducial model, which does not account for destruction of \Lya photons by dust, indicates that the \sblya is higher for galaxies with higher stellar mass out to large distances. When including dust treatment \citep[see appendix A4 of][]{byrohl/etal:2021}, the correlation between stellar mass and outer \sblya level weakens because massive galaxies are more strongly affected by dust attenuation. The resulting \sblya profiles are more similar across stellar masses, better matching our findings.

\subsection{Comparison with Previous Results}
\label{subsec:comparison_with_previous_results}

Figure \ref{fig:literature_comparison} compares the surface luminosity profile of \oiii galaxies and that of the LAEs with previous results for LAEs and LBGs at redshifts $2<z<4$. We show the profiles as a function of physical distance because most of the data lies within the virial radii of the galaxies.
The profiles that were given as a function of comoving distance or did not contain the $(1+z)^4$ factor were adjusted using one redshift for each sample (see caption of Figure \ref{fig:literature_comparison}).
Because of the smaller PSF of MUSE, the profiles of \citet{wisotzki/etal:2018} and \citet{kusakabe/etal:2022} are steeper in the core than our profiles. We therefore show the convolved profiles with the PSF model following the stacked star profile in the \oiii galaxy observations (Moffat function with $\beta=2.2$ and FWHM$=1.47$ convolved with the VIRUS fiber profile), as though VIRUS observed these profiles at $z=2.1$.
Despite differences at small distances, all LAE and LBG profiles are similar at intermediate distances ($20\,\mathrm{kpc}\lesssim r \lesssim 80\,\mathrm{kpc}$).

\subsection{Emission Mechanism}
\subsubsection{Star formation in other galaxies}
To find out whether the measured \Lya emission can be powered by star formation alone, we estimate the required star formation rate density (SFRD) from the measured \Lya luminosity within $800\,\rm kpc$ of the \oiii sample ($L_{\rm Ly\alpha} = (2.3\pm1.3)\times 10^{43}\,\rm erg \,s^{-1}$).
Using \mbox{$L_{\rm Ly\alpha} = 10^{42}\,{\rm erg\, s^{-1}}\times {\rm SFR}\,{\rm M_\odot^{-1} \, yr}$} \citep[see][]{Dijkstra:2019}, we find ${\rm SFRD}=0.05\pm 0.03\,{\rm M_\odot \, yr^{-1}\,cMpc^{-3}}$. This value is smaller than those in the literature \citep[$\simeq 0.1\, M_\odot\,\mathrm{yr}^{-1}\,\mathrm{cMpc}^{-3}$; summarized by][]{rowan-robinson/etal:2016}, implying that star-formation-induced photons can account for the \Lya emission out to $800$~kpc.

\subsubsection{Star formation in the central galaxy}
The median SFR of the \oiii galaxies is $10^{0.960}\,M_\odot\,\mathrm{yr}^{-1}$. Using the same $L_{\rm Ly\alpha}$-SFR relation as above, we expect an intrinsic \Lya luminosity $L^{\rm int}_{\rm Ly\alpha} \simeq 9.1\times10^{42}\,\rm erg\, s^{-1}$. Because this is consistent with the measured luminosity, the \Lya photons could originate from the central galaxies if the escape fraction is close to one.
This scenario is disfavoured because of the small measured escape fraction of $6^{+0.6}_{-0.5}\%$ \citep{weiss2021}, as well as the theoretically expected one. Using the standard relation for the optical depth due to dust extinction of \citet{calzetti/etal:2000} and \citet{verhamme/etal:2006} yields $f^{\rm Ly\alpha}_{\rm esc}\leq\exp\left\{-\tau_{\rm dust}^{\rm Ly\alpha}\right\}\approx 0.21$ for the median E(B-V) of our sample.

\subsubsection{Fluorescence}
\citet{cantalupo/etal:2005} predict a \Lya surface brightness through fluorescence from the UV background of $3.67\times 10^{-20}\,\sbunits$ at $z\sim3$. The photoionization rate of the UV background changes little from $z=3$ to $z=2$ \citep[][]{fauchergiguere/etal:2020}. Accounting for cosmic dimming, this value would be $\simeq 10^{-19}\,\sbunits$ at $z=2.1$, which is consistent with the intermediate and outer points of the radial profiles, but too low to explain levels at small distances.

\subsubsection{Cooling radiation}
\Lya photons can be emitted through collisional excitation and recombination in cooling gas flowing into a galaxy. 
The subsequent scattering in an inflowing medium can lead to a blue-shift of the \Lya line \citep[][]{dijkstra/haiman/spaans:2006}. While the scattering and blue-shift may be negligible due to the low volume-filling factor of cold streams, we expect a filamentary morphology of the \Lya emission \citep[][]{,dijkstra/loeb:2009}.
We cannot test whether the \Lya line is blue-shifted because of the low spectral resolution and the high redshift uncertainty of the \oiii galaxies. Detecting the filamentary structure requires deep observations of individual \Lya halos rather than stacking and circular averaging.

\section{Summary}\label{sec:summary}
We measure the \Lya emission
out to $800\,\mathrm{kpc}$ around \numgal [\ion{O}{3}]-selected galaxies at $1.9<z<2.35$.
While the central surface brightness in the core ($r<10\,\mathrm{kpc}$) is fainter than that of the median redshift-adjusted \sblya profile of 968 LAEs at $1.9<z<3.5$ by an order of magnitude,
the \Lya surface brightness in the outer parts ($r>40$~kpc) reaches the same surface brightness as that of the LAEs.

This result supports the picture in which photons originating from outside of the central galaxies dominate the \sblya profiles at large radii.
These photons either originate from other dark matter halos or satellite galaxies, or are emitted through fluorescence or cooling radiation in the CGM. While we cannot exclude any of these sources, star formation alone can account for the integrated \Lya emission out to $800$~kpc, and fluorescence from the UV background is sufficient to explain the surface brightness at intermediate distances.

\begin{acknowledgments}
We thank C. Byrohl for helpful and interesting discussions and F. Arrigoni Battaia, V. Gonz\'alez Lobos, and C. Peroux for comments on the draft. We also thank the anonymous referee for the helpful review.
EK's work was supported in part by the Deutsche Forschungsgemeinschaft (DFG, German Research Foundation) under Germany's Excellence Strategy - EXC-2094 - 390783311.

HETDEX is led by the University of Texas at Austin, McDonald Observatory, and Department of Astronomy with participation from the Ludwig-Maximilians-Universit\"at M\"unchen, Max-Planck-Institut f\"ur Extraterrestrische Physik (MPE), Leibniz-Institut f\"ur Astrophysik Potsdam (AIP), Texas A\&M University, Pennsylvania State University, Institut f\"ur Astrophysik G\"ottingen, The University of Oxford, Max-Planck-Institut f\"ur Astrophysik (MPA), The University of Tokyo, and Missouri University of Science and Technology. In addition to Institutional support, HETDEX is funded by the National Science Foundation (grant AST-0926815), the State of Texas, the US Air Force (AFRL FA9451-04-2-0355), and generous support from private individuals and foundations. 

The observations were obtained with the Hobby-Eberly Telescope (HET), which is a joint project of the University of Texas at Austin, the Pennsylvania State University, Ludwig-Maximilians-Universität München, and Georg-August-Universität Göttingen. The HET is named in honor of its principal benefactors, William P. Hobby and Robert E. Eberly. 

VIRUS is a joint project of the University of Texas at Austin, Leibniz-Institut f\"ur Astrophysik Potsdam (AIP), Texas A\&M University (TAMU), Max-Planck-Institut f\"ur Extraterrestrische Physik (MPE), Ludwig-Maximilians-Universit\"at M\"unchen, Pennsylvania State University, Institut f\"ur Astrophysik G\"ottingen, University of Oxford, Max-Planck-Institut f\"ur Astrophysik (MPA), and The University of Tokyo. 

The authors acknowledge the Texas Advanced Computing Center (TACC) at The University of Texas at Austin for providing high performance computing, visualization, and storage resources that have contributed to the research results reported within this paper. URL: http://www.tacc.utexas.edu.

This work is based on observations taken by the 3D-HST Treasury Program (GO 12177 and 12328) with the NASA/ESA HST, which is operated by the Association of Universities for Research in Astronomy, Inc., under NASA contract NAS5-26555.

The Institute for Gravitation and the Cosmos is supported by the Eberly College of Science and the Office of the Senior Vice President for Research at the Pennsylvania State University. 

This research made use of NASA's Astrophysics Data System Bibliographic Services.

\end{acknowledgments}

\software{Astropy \citep{astropy:2018},
        Numpy
        \citep{harris2020array},
        Scipy
        \citep{2020SciPy-NMeth},
        Matplotlib
        \citep{Hunter:2007}}

\bibliography{elg}{}

\begin{thebibliography}{}
\expandafter\ifx\csname natexlab\endcsname\relax\def\natexlab#1{#1}\fi
\providecommand{\url}[1]{\href{#1}{#1}}
\providecommand{\dodoi}[1]{doi:~\href{http://doi.org/#1}{\nolinkurl{#1}}}
\providecommand{\doeprint}[1]{\href{http://ascl.net/#1}{\nolinkurl{http://ascl.net/#1}}}
\providecommand{\doarXiv}[1]{\href{https://arxiv.org/abs/#1}{\nolinkurl{https://arxiv.org/abs/#1}}}

\bibitem[{{Astropy Collaboration} {et~al.}(2018){Astropy Collaboration},
  {Price-Whelan}, {Sip{H{o}}cz}, {G{"u}nther}, {Lim}, {Crawford}, {Conseil},
  {Shupe}, {Craig}, {Dencheva}, {Ginsburg}, {Vand erPlas}, {Bradley},
  {P{'e}rez-Su{'a}rez}, {de Val-Borro}, {Aldcroft}, {Cruz}, {Robitaille},
  {Tollerud}, {Ardelean}, {Babej}, {Bach}, {Bachetti}, {Bakanov}, {Bamford},
  {Barentsen}, {Barmby}, {Baumbach}, {Berry}, {Biscani}, {Boquien}, {Bostroem},
  {Bouma}, {Brammer}, {Bray}, {Breytenbach}, {Buddelmeijer}, {Burke},
  {Calderone}, {Cano Rodr{'i}guez}, {Cara}, {Cardoso}, {Cheedella}, {Copin},
  {Corrales}, {Crichton}, {D'Avella}, {Deil}, {Depagne}, {Dietrich}, {Donath},
  {Droettboom}, {Earl}, {Erben}, {Fabbro}, {Ferreira}, {Finethy}, {Fox},
  {Garrison}, {Gibbons}, {Goldstein}, {Gommers}, {Greco}, {Greenfield},
  {Groener}, {Grollier}, {Hagen}, {Hirst}, {Homeier}, {Horton}, {Hosseinzadeh},
  {Hu}, {Hunkeler}, {Ivezi{'c}}, {Jain}, {Jenness}, {Kanarek}, {Kendrew},
  {Kern}, {Kerzendorf}, {Khvalko}, {King}, {Kirkby}, {Kulkarni}, {Kumar},
  {Lee}, {Lenz}, {Littlefair}, {Ma}, {Macleod}, {Mastropietro}, {McCully},
  {Montagnac}, {Morris}, {Mueller}, {Mumford}, {Muna}, {Murphy}, {Nelson},
  {Nguyen}, {Ninan}, {N{"o}the}, {Ogaz}, {Oh}, {Parejko}, {Parley}, {Pascual},
  {Patil}, {Patil}, {Plunkett}, {Prochaska}, {Rastogi}, {Reddy Janga},
  {Sabater}, {Sakurikar}, {Seifert}, {Sherbert}, {Sherwood-Taylor}, {Shih},
  {Sick}, {Silbiger}, {Singanamalla}, {Singer}, {Sladen}, {Sooley},
  {Sornarajah}, {Streicher}, {Teuben}, {Thomas}, {Tremblay}, {Turner},
  {Terr{'o}n}, {van Kerkwijk}, {de la Vega}, {Watkins}, {Weaver}, {Whitmore},
  {Woillez}, {Zabalza}, \& {Astropy Contributors}}]{astropy:2018}
{Astropy Collaboration}, {Price-Whelan}, A.~M., {Sip{H{o}}cz}, B.~M., {et~al.}
  2018, \aj, 156, 123, \dodoi{10.3847/1538-3881/aabc4f}

\bibitem[{{Behroozi} {et~al.}(2019){Behroozi}, {Wechsler}, {Hearin}, \&
  {Conroy}}]{behroozi/etal:2019}
{Behroozi}, P., {Wechsler}, R.~H., {Hearin}, A.~P., \& {Conroy}, C. 2019,
  \mnras, 488, 3143, \dodoi{10.1093/mnras/stz1182}

\bibitem[{{Bowman} {et~al.}(2019){Bowman}, {Zeimann}, {Ciardullo}, {Gronwall},
  {Schneider}, {McCarron}, {Weiss}, {Yang}, \& {Hagen}}]{bowman2019}
{Bowman}, W.~P., {Zeimann}, G.~R., {Ciardullo}, R., {et~al.} 2019, \apj, 875,
  152, \dodoi{10.3847/1538-4357/ab108a}

\bibitem[{{Bowman} {et~al.}(2020){Bowman}, {Zeimann}, {Nagaraj}, {Ciardullo},
  {Gronwall}, {McCarron}, {Weiss}, {Molina}, {Belles}, \&
  {Schneider}}]{bowman2020}
{Bowman}, W.~P., {Zeimann}, G.~R., {Nagaraj}, G., {et~al.} 2020, \apj, 899, 7,
  \dodoi{10.3847/1538-4357/ab9f3c}

\bibitem[{{Bowman} {et~al.}(2021){Bowman}, {Ciardullo}, {Zeimann}, {Gronwall},
  {Jeong}, {Nagaraj}, {Abelson}, {Weiss}, {Molina}, \&
  {Schneider}}]{bowman/etal:2021}
{Bowman}, W.~P., {Ciardullo}, R., {Zeimann}, G.~R., {et~al.} 2021, \apj, 920,
  78, \dodoi{10.3847/1538-4357/ac1a0e}

\bibitem[{{Brammer} {et~al.}(2012){Brammer}, {van Dokkum}, {Franx},
  {Fumagalli}, {Patel}, {Rix}, {Skelton}, {Kriek}, {Nelson}, {Schmidt},
  {Bezanson}, {da Cunha}, {Erb}, {Fan}, {F{\"o}rster Schreiber}, {Illingworth},
  {Labb{\'e}}, {Leja}, {Lundgren}, {Magee}, {Marchesini}, {McCarthy},
  {Momcheva}, {Muzzin}, {Quadri}, {Steidel}, {Tal}, {Wake}, {Whitaker}, \&
  {Williams}}]{brammer2012}
{Brammer}, G.~B., {van Dokkum}, P.~G., {Franx}, M., {et~al.} 2012, \apjs, 200,
  13, \dodoi{10.1088/0067-0049/200/2/13}

\bibitem[{{Bryan} \& {Norman}(1998)}]{bryan/norman:1998}
{Bryan}, G.~L., \& {Norman}, M.~L. 1998, \apj, 495, 80, \dodoi{10.1086/305262}

\bibitem[{{Byrohl} {et~al.}(2021){Byrohl}, {Nelson}, {Behrens}, {Kostyuk},
  {Glatzle}, {Pillepich}, {Hernquist}, {Marinacci}, \&
  {Vogelsberger}}]{byrohl/etal:2021}
{Byrohl}, C., {Nelson}, D., {Behrens}, C., {et~al.} 2021, \mnras, 506, 5129,
  \dodoi{10.1093/mnras/stab1958}

\bibitem[{{Calzetti} {et~al.}(2000){Calzetti}, {Armus}, {Bohlin}, {Kinney},
  {Koornneef}, \& {Storchi-Bergmann}}]{calzetti/etal:2000}
{Calzetti}, D., {Armus}, L., {Bohlin}, R.~C., {et~al.} 2000, \apj, 533, 682,
  \dodoi{10.1086/308692}

\bibitem[{{Cantalupo} {et~al.}(2005){Cantalupo}, {Porciani}, {Lilly}, \&
  {Miniati}}]{cantalupo/etal:2005}
{Cantalupo}, S., {Porciani}, C., {Lilly}, S.~J., \& {Miniati}, F. 2005, \apj,
  628, 61, \dodoi{10.1086/430758}

\bibitem[{{Dijkstra}(2019)}]{Dijkstra:2019}
{Dijkstra}, M. 2019, Saas-Fee Advanced Course, 46, 1,
  \dodoi{10.1007/978-3-662-59623-4_1}

\bibitem[{{Dijkstra} {et~al.}(2006){Dijkstra}, {Haiman}, \&
  {Spaans}}]{dijkstra/haiman/spaans:2006}
{Dijkstra}, M., {Haiman}, Z., \& {Spaans}, M. 2006, \apj, 649, 14,
  \dodoi{10.1086/506243}

\bibitem[{{Dijkstra} \& {Loeb}(2009)}]{dijkstra/loeb:2009}
{Dijkstra}, M., \& {Loeb}, A. 2009, \mnras, 400, 1109,
  \dodoi{10.1111/j.1365-2966.2009.15533.x}

\bibitem[{{Erb} {et~al.}(2016){Erb}, {Pettini}, {Steidel}, {Strom}, {Rudie},
  {Trainor}, {Shapley}, \& {Reddy}}]{erb2016}
{Erb}, D.~K., {Pettini}, M., {Steidel}, C.~C., {et~al.} 2016, \apj, 830, 52,
  \dodoi{10.3847/0004-637X/830/1/52}

\bibitem[{{Faucher-Gigu{\`e}re}(2020)}]{fauchergiguere/etal:2020}
{Faucher-Gigu{\`e}re}, C.-A. 2020, \mnras, 493, 1614,
  \dodoi{10.1093/mnras/staa302}

\bibitem[{{Gaia Collaboration} {et~al.}(2018){Gaia Collaboration}, {Brown},
  {Vallenari}, {Prusti}, {de Bruijne}, {Babusiaux}, {Bailer-Jones}, {Biermann},
  {Evans}, {Eyer}, {Jansen}, {Jordi}, {Klioner}, {Lammers}, {Lindegren},
  {Luri}, {Mignard}, {Panem}, {Pourbaix}, {Randich}, {Sartoretti}, {Siddiqui},
  {Soubiran}, {van Leeuwen}, {Walton}, {Arenou}, {Bastian}, {Cropper},
  {Drimmel}, {Katz}, {Lattanzi}, {Bakker}, {Cacciari}, {Casta{\~n}eda},
  {Chaoul}, {Cheek}, {De Angeli}, {Fabricius}, {Guerra}, {Holl}, {Masana},
  {Messineo}, {Mowlavi}, {Nienartowicz}, {Panuzzo}, {Portell}, {Riello},
  {Seabroke}, {Tanga}, {Th{\'e}venin}, {Gracia-Abril}, {Comoretto},
  {Garcia-Reinaldos}, {Teyssier}, {Altmann}, {Andrae}, {Audard},
  {Bellas-Velidis}, {Benson}, {Berthier}, {Blomme}, {Burgess}, {Busso},
  {Carry}, {Cellino}, {Clementini}, {Clotet}, {Creevey}, {Davidson}, {De
  Ridder}, {Delchambre}, {Dell'Oro}, {Ducourant},
  {Fern{\'a}ndez-Hern{\'a}ndez}, {Fouesneau}, {Fr{\'e}mat}, {Galluccio},
  {Garc{\'\i}a-Torres}, {Gonz{\'a}lez-N{\'u}{\~n}ez}, {Gonz{\'a}lez-Vidal},
  {Gosset}, {Guy}, {Halbwachs}, {Hambly}, {Harrison}, {Hern{\'a}ndez},
  {Hestroffer}, {Hodgkin}, {Hutton}, {Jasniewicz}, {Jean-Antoine-Piccolo},
  {Jordan}, {Korn}, {Krone-Martins}, {Lanzafame}, {Lebzelter}, {L{\"o}ffler},
  {Manteiga}, {Marrese}, {Mart{\'\i}n-Fleitas}, {Moitinho}, {Mora}, {Muinonen},
  {Osinde}, {Pancino}, {Pauwels}, {Petit}, {Recio-Blanco}, {Richards},
  {Rimoldini}, {Robin}, {Sarro}, {Siopis}, {Smith}, {Sozzetti}, {S{\"u}veges},
  {Torra}, {van Reeven}, {Abbas}, {Abreu Aramburu}, {Accart}, {Aerts},
  {Altavilla}, {{\'A}lvarez}, {Alvarez}, {Alves}, {Anderson}, {Andrei},
  {Anglada Varela}, {Antiche}, {Antoja}, {Arcay}, {Astraatmadja}, {Bach},
  {Baker}, {Balaguer-N{\'u}{\~n}ez}, {Balm}, {Barache}, {Barata}, {Barbato},
  {Barblan}, {Barklem}, {Barrado}, {Barros}, {Barstow}, {Bartholom{\'e}
  Mu{\~n}oz}, {Bassilana}, {Becciani}, {Bellazzini}, {Berihuete}, {Bertone},
  {Bianchi}, {Bienaym{\'e}}, {Blanco-Cuaresma}, {Boch}, {Boeche}, {Bombrun},
  {Borrachero}, {Bossini}, {Bouquillon}, {Bourda}, {Bragaglia}, {Bramante},
  {Breddels}, {Bressan}, {Brouillet}, {Br{\"u}semeister}, {Brugaletta},
  {Bucciarelli}, {Burlacu}, {Busonero}, {Butkevich}, {Buzzi}, {Caffau},
  {Cancelliere}, {Cannizzaro}, {Cantat-Gaudin}, {Carballo}, {Carlucci},
  {Carrasco}, {Casamiquela}, {Castellani}, {Castro-Ginard}, {Charlot},
  {Chemin}, {Chiavassa}, {Cocozza}, {Costigan}, {Cowell}, {Crifo}, {Crosta},
  {Crowley}, {Cuypers}, {Dafonte}, {Damerdji}, {Dapergolas}, {David}, {David},
  {de Laverny}, {De Luise}, {De March}, {de Martino}, {de Souza}, {de Torres},
  {Debosscher}, {del Pozo}, {Delbo}, {Delgado}, {Delgado}, {Di Matteo},
  {Diakite}, {Diener}, {Distefano}, {Dolding}, {Drazinos}, {Dur{\'a}n},
  {Edvardsson}, {Enke}, {Eriksson}, {Esquej}, {Eynard Bontemps}, {Fabre},
  {Fabrizio}, {Faigler}, {Falc{\~a}o}, {Farr{\`a}s Casas}, {Federici},
  {Fedorets}, {Fernique}, {Figueras}, {Filippi}, {Findeisen}, {Fonti},
  {Fraile}, {Fraser}, {Fr{\'e}zouls}, {Gai}, {Galleti}, {Garabato},
  {Garc{\'\i}a-Sedano}, {Garofalo}, {Garralda}, {Gavel}, {Gavras}, {Gerssen},
  {Geyer}, {Giacobbe}, {Gilmore}, {Girona}, {Giuffrida}, {Glass}, {Gomes},
  {Granvik}, {Gueguen}, {Guerrier}, {Guiraud}, {Guti{\'e}rrez-S{\'a}nchez},
  {Haigron}, {Hatzidimitriou}, {Hauser}, {Haywood}, {Heiter}, {Helmi}, {Heu},
  {Hilger}, {Hobbs}, {Hofmann}, {Holland}, {Huckle}, {Hypki}, {Icardi},
  {Jan{\ss}en}, {Jevardat de Fombelle}, {Jonker}, {Juh{\'a}sz}, {Julbe},
  {Karampelas}, {Kewley}, {Klar}, {Kochoska}, {Kohley}, {Kolenberg},
  {Kontizas}, {Kontizas}, {Koposov}, {Kordopatis}, {Kostrzewa-Rutkowska},
  {Koubsky}, {Lambert}, {Lanza}, {Lasne}, {Lavigne}, {Le Fustec}, {Le
  Poncin-Lafitte}, {Lebreton}, {Leccia}, {Leclerc}, {Lecoeur-Taibi},
  {Lenhardt}, {Leroux}, {Liao}, {Licata}, {Lindstr{\o}m}, {Lister}, {Livanou},
  {Lobel}, {L{\'o}pez}, {Managau}, {Mann}, {Mantelet}, {Marchal}, {Marchant},
  {Marconi}, {Marinoni}, {Marschalk{\'o}}, {Marshall}, {Martino}, {Marton},
  {Mary}, {Massari}, {Matijevi{\v{c}}}, {Mazeh}, {McMillan}, {Messina},
  {Michalik}, {Millar}, {Molina}, {Molinaro}, {Moln{\'a}r}, {Montegriffo},
  {Mor}, {Morbidelli}, {Morel}, {Morris}, {Mulone}, {Muraveva}, {Musella},
  {Nelemans}, {Nicastro}, {Noval}, {O'Mullane}, {Ord{\'e}novic},
  {Ord{\'o}{\~n}ez-Blanco}, {Osborne}, {Pagani}, {Pagano}, {Pailler},
  {Palacin}, {Palaversa}, {Panahi}, {Pawlak}, {Piersimoni}, {Pineau}, {Plachy},
  {Plum}, {Poggio}, {Poujoulet}, {Pr{\v{s}}a}, {Pulone}, {Racero}, {Ragaini},
  {Rambaux}, {Ramos-Lerate}, {Regibo}, {Reyl{\'e}}, {Riclet}, {Ripepi}, {Riva},
  {Rivard}, {Rixon}, {Roegiers}, {Roelens}, {Romero-G{\'o}mez}, {Rowell},
  {Royer}, {Ruiz-Dern}, {Sadowski}, {Sagrist{\`a} Sell{\'e}s}, {Sahlmann},
  {Salgado}, {Salguero}, {Sanna}, {Santana-Ros}, {Sarasso}, {Savietto},
  {Schultheis}, {Sciacca}, {Segol}, {Segovia}, {S{\'e}gransan}, {Shih},
  {Siltala}, {Silva}, {Smart}, {Smith}, {Solano}, {Solitro}, {Sordo}, {Soria
  Nieto}, {Souchay}, {Spagna}, {Spoto}, {Stampa}, {Steele},
  {Steidelm{\"u}ller}, {Stephenson}, {Stoev}, {Suess}, {Surdej}, {Szabados},
  {Szegedi-Elek}, {Tapiador}, {Taris}, {Tauran}, {Taylor}, {Teixeira},
  {Terrett}, {Teyssandier}, {Thuillot}, {Titarenko}, {Torra Clotet}, {Turon},
  {Ulla}, {Utrilla}, {Uzzi}, {Vaillant}, {Valentini}, {Valette}, {van Elteren},
  {Van Hemelryck}, {van Leeuwen}, {Vaschetto}, {Vecchiato}, {Veljanoski},
  {Viala}, {Vicente}, {Vogt}, {von Essen}, {Voss}, {Votruba}, {Voutsinas},
  {Walmsley}, {Weiler}, {Wertz}, {Wevers}, {Wyrzykowski}, {Yoldas},
  {{\v{Z}}erjal}, {Ziaeepour}, {Zorec}, {Zschocke}, {Zucker}, {Zurbach}, \&
  {Zwitter}}]{gaiadr2/2018}
{Gaia Collaboration}, {Brown}, A.~G.~A., {Vallenari}, A., {et~al.} 2018, \aap,
  616, A1, \dodoi{10.1051/0004-6361/201833051}

\bibitem[{{Gebhardt} {et~al.}(2021){Gebhardt}, {Mentuch Cooper}, {Ciardullo},
  {Acquaviva}, {Bender}, {Bowman}, {Castanheira}, {Dalton}, {Davis}, {de Jong},
  {DePoy}, {Devarakonda}, {Dongsheng}, {Drory}, {Fabricius}, {Farrow},
  {Feldmeier}, {Finkelstein}, {Froning}, {Gawiser}, {Gronwall}, {Herold},
  {Hill}, {Hopp}, {House}, {Janowiecki}, {Jarvis}, {Jeong}, {Jogee}, {Kakuma},
  {Kelz}, {Kollatschny}, {Komatsu}, {Krumpe}, {Landriau}, {Liu}, {Niemeyer},
  {MacQueen}, {Marshall}, {Mawatari}, {McLinden}, {Mukae}, {Nagaraj}, {Ono},
  {Ouchi}, {Papovich}, {Sakai}, {Saito}, {Schneider}, {Schulze},
  {Shanmugasundararaj}, {Shetrone}, {Sneden}, {Snigula}, {Steinmetz}, {Thomas},
  {Thomas}, {Tuttle}, {Urrutia}, {Wisotzki}, {Wold}, {Zeimann}, \&
  {Zhang}}]{gebhardt/etal:2021}
{Gebhardt}, K., {Mentuch Cooper}, E., {Ciardullo}, R., {et~al.} 2021, \apj,
  923, 217, \dodoi{10.3847/1538-4357/ac2e03}

\bibitem[{{Gould} \& {Weinberg}(1996)}]{gould/weinberg:1996}
{Gould}, A., \& {Weinberg}, D.~H. 1996, \apj, 468, 462, \dodoi{10.1086/177707}

\bibitem[{{Grogin} {et~al.}(2011){Grogin}, {Kocevski}, {Faber}, {Ferguson},
  {Koekemoer}, {Riess}, {Acquaviva}, {Alexander}, {Almaini}, {Ashby}, {Barden},
  {Bell}, {Bournaud}, {Brown}, {Caputi}, {Casertano}, {Cassata}, {Castellano},
  {Challis}, {Chary}, {Cheung}, {Cirasuolo}, {Conselice}, {Roshan Cooray},
  {Croton}, {Daddi}, {Dahlen}, {Dav{\'e}}, {de Mello}, {Dekel}, {Dickinson},
  {Dolch}, {Donley}, {Dunlop}, {Dutton}, {Elbaz}, {Fazio}, {Filippenko},
  {Finkelstein}, {Fontana}, {Gardner}, {Garnavich}, {Gawiser}, {Giavalisco},
  {Grazian}, {Guo}, {Hathi}, {H{\"a}ussler}, {Hopkins}, {Huang}, {Huang},
  {Jha}, {Kartaltepe}, {Kirshner}, {Koo}, {Lai}, {Lee}, {Li}, {Lotz}, {Lucas},
  {Madau}, {McCarthy}, {McGrath}, {McIntosh}, {McLure}, {Mobasher},
  {Moustakas}, {Mozena}, {Nandra}, {Newman}, {Niemi}, {Noeske}, {Papovich},
  {Pentericci}, {Pope}, {Primack}, {Rajan}, {Ravindranath}, {Reddy}, {Renzini},
  {Rix}, {Robaina}, {Rodney}, {Rosario}, {Rosati}, {Salimbeni}, {Scarlata},
  {Siana}, {Simard}, {Smidt}, {Somerville}, {Spinrad}, {Straughn}, {Strolger},
  {Telford}, {Teplitz}, {Trump}, {van der Wel}, {Villforth}, {Wechsler},
  {Weiner}, {Wiklind}, {Wild}, {Wilson}, {Wuyts}, {Yan}, \& {Yun}}]{grogin2011}
{Grogin}, N.~A., {Kocevski}, D.~D., {Faber}, S.~M., {et~al.} 2011, \apjs, 197,
  35, \dodoi{10.1088/0067-0049/197/2/35}

\bibitem[{{Hagen} {et~al.}(2016){Hagen}, {Zeimann}, {Behrens}, {Ciardullo},
  {Grasshorn Gebhardt}, {Gronwall}, {Bridge}, {Fox}, {Schneider}, {Trump},
  {Blanc}, {Chiang}, {Chonis}, {Finkelstein}, {Hill}, {Jogee}, \&
  {Gawiser}}]{hagen/etal:2016}
{Hagen}, A., {Zeimann}, G.~R., {Behrens}, C., {et~al.} 2016, \apj, 817, 79,
  \dodoi{10.3847/0004-637X/817/1/79}

\bibitem[{{Haiman} {et~al.}(2000){Haiman}, {Spaans}, \&
  {Quataert}}]{haiman/spaans/quataert:2000}
{Haiman}, Z., {Spaans}, M., \& {Quataert}, E. 2000, \apjl, 537, L5,
  \dodoi{10.1086/312754}

\bibitem[{Harris {et~al.}(2020)Harris, Millman, van~der Walt, Gommers,
  Virtanen, Cournapeau, Wieser, Taylor, Berg, Smith, Kern, Picus, Hoyer, van
  Kerkwijk, Brett, Haldane, del R{\'{i}}o, Wiebe, Peterson,
  G{\'{e}}rard-Marchant, Sheppard, Reddy, Weckesser, Abbasi, Gohlke, \&
  Oliphant}]{harris2020array}
Harris, C.~R., Millman, K.~J., van~der Walt, S.~J., {et~al.} 2020, Nature, 585,
  357, \dodoi{10.1038/s41586-020-2649-2}

\bibitem[{{Hathi} {et~al.}(2016){Hathi}, {Le F{\`e}vre}, {Ilbert}, {Cassata},
  {Tasca}, {Lemaux}, {Garilli}, {Le Brun}, {Maccagni}, {Pentericci}, {Thomas},
  {Vanzella}, {Zamorani}, {Zucca}, {Amor{\'\i}n}, {Bardelli}, {Cassar{\`a}},
  {Castellano}, {Cimatti}, {Cucciati}, {Durkalec}, {Fontana}, {Giavalisco},
  {Grazian}, {Guaita}, {Koekemoer}, {Paltani}, {Pforr}, {Ribeiro}, {Schaerer},
  {Scodeggio}, {Sommariva}, {Talia}, {Tresse}, {Vergani}, {Capak}, {Charlot},
  {Contini}, {Cuby}, {de la Torre}, {Dunlop}, {Fotopoulou},
  {L{\'o}pez-Sanjuan}, {Mellier}, {Salvato}, {Scoville}, {Taniguchi}, \&
  {Wang}}]{hathi/etal:2016}
{Hathi}, N.~P., {Le F{\`e}vre}, O., {Ilbert}, O., {et~al.} 2016, \aap, 588,
  A26, \dodoi{10.1051/0004-6361/201526012}

\bibitem[{{Hill} {et~al.}(2021){Hill}, {Lee}, {MacQueen}, {Kelz}, {Drory},
  {Vattiat}, {Good}, {Ramsey}, {Kriel}, {Peterson}, {DePoy}, {Gebhardt},
  {Marshall}, {Tuttle}, {Bauer}, {Chonis}, {Fabricius}, {Froning},
  {H{\"a}user}, {Indahl}, {Jahn}, {Landriau}, {Leck}, {Montesano}, {Prochaska},
  {Snigula}, {Zeimann}, {Bryant}, {Damm}, {Fowler}, {Janowiecki}, {Martin},
  {Mrozinski}, {Odewahn}, {Rostopchin}, {Shetrone}, {Spencer}, {Mentuch
  Cooper}, {Armandroff}, {Bender}, {Dalton}, {Hopp}, {Komatsu}, {Nicklas},
  {Ramsey}, {Roth}, {Schneider}, {Sneden}, \& {Steinmetz}}]{hill/etal:2021}
{Hill}, G.~J., {Lee}, H., {MacQueen}, P.~J., {et~al.} 2021, \aj, 162, 298,
  \dodoi{10.3847/1538-3881/ac2c02}

\bibitem[{Hunter(2007)}]{Hunter:2007}
Hunter, J.~D. 2007, Computing in Science \& Engineering, 9, 90,
  \dodoi{10.1109/MCSE.2007.55}

\bibitem[{{Kikuchihara} {et~al.}(2021){Kikuchihara}, {Harikane}, {Ouchi},
  {Ono}, {Shibuya}, {Itoh}, {Kakuma}, {Inoue}, {Kusakabe}, {Shimasaku},
  {Momose}, {Sugahara}, {Kikuta}, {Saito}, {Kashikawa}, {Zhang}, \&
  {Lee}}]{kikuchihara/etal:2021}
{Kikuchihara}, S., {Harikane}, Y., {Ouchi}, M., {et~al.} 2021, arXiv e-prints,
  arXiv:2108.09288.
\newblock \doarXiv{2108.09288}

\bibitem[{{Koekemoer} {et~al.}(2011){Koekemoer}, {Faber}, {Ferguson}, {Grogin},
  {Kocevski}, {Koo}, {Lai}, {Lotz}, {Lucas}, {McGrath}, {Ogaz}, {Rajan},
  {Riess}, {Rodney}, {Strolger}, {Casertano}, {Castellano}, {Dahlen},
  {Dickinson}, {Dolch}, {Fontana}, {Giavalisco}, {Grazian}, {Guo}, {Hathi},
  {Huang}, {van der Wel}, {Yan}, {Acquaviva}, {Alexander}, {Almaini}, {Ashby},
  {Barden}, {Bell}, {Bournaud}, {Brown}, {Caputi}, {Cassata}, {Challis},
  {Chary}, {Cheung}, {Cirasuolo}, {Conselice}, {Roshan Cooray}, {Croton},
  {Daddi}, {Dav{\'e}}, {de Mello}, {de Ravel}, {Dekel}, {Donley}, {Dunlop},
  {Dutton}, {Elbaz}, {Fazio}, {Filippenko}, {Finkelstein}, {Frazer}, {Gardner},
  {Garnavich}, {Gawiser}, {Gruetzbauch}, {Hartley}, {H{\"a}ussler},
  {Herrington}, {Hopkins}, {Huang}, {Jha}, {Johnson}, {Kartaltepe},
  {Khostovan}, {Kirshner}, {Lani}, {Lee}, {Li}, {Madau}, {McCarthy},
  {McIntosh}, {McLure}, {McPartland}, {Mobasher}, {Moreira}, {Mortlock},
  {Moustakas}, {Mozena}, {Nandra}, {Newman}, {Nielsen}, {Niemi}, {Noeske},
  {Papovich}, {Pentericci}, {Pope}, {Primack}, {Ravindranath}, {Reddy},
  {Renzini}, {Rix}, {Robaina}, {Rosario}, {Rosati}, {Salimbeni}, {Scarlata},
  {Siana}, {Simard}, {Smidt}, {Snyder}, {Somerville}, {Spinrad}, {Straughn},
  {Telford}, {Teplitz}, {Trump}, {Vargas}, {Villforth}, {Wagner}, {Wandro},
  {Wechsler}, {Weiner}, {Wiklind}, {Wild}, {Wilson}, {Wuyts}, \&
  {Yun}}]{koekemoer2011}
{Koekemoer}, A.~M., {Faber}, S.~M., {Ferguson}, H.~C., {et~al.} 2011, \apjs,
  197, 36, \dodoi{10.1088/0067-0049/197/2/36}

\bibitem[{{Kusakabe} {et~al.}(2022){Kusakabe}, {Verhamme}, {Blaizot}, {Garel},
  {Wisotzki}, {Leclercq}, {Bacon}, {Schaye}, {Gallego}, {Kerutt}, {Matthee},
  {Maseda}, {Nanayakkara}, {Pello}, {Richard}, {Tresse}, {Urrutia}, \&
  {Vitte}}]{kusakabe/etal:2022}
{Kusakabe}, H., {Verhamme}, A., {Blaizot}, J., {et~al.} 2022, arXiv e-prints,
  arXiv:2201.07257.
\newblock \doarXiv{2201.07257}

\bibitem[{{Leclercq} {et~al.}(2017){Leclercq}, {Bacon}, {Wisotzki}, {Mitchell},
  {Garel}, {Verhamme}, {Blaizot}, {Hashimoto}, {Herenz}, {Conseil},
  {Cantalupo}, {Inami}, {Contini}, {Richard}, {Maseda}, {Schaye}, {Marino},
  {Akhlaghi}, {Brinchmann}, \& {Carollo}}]{leclercq/etal:2017}
{Leclercq}, F., {Bacon}, R., {Wisotzki}, L., {et~al.} 2017, \aap, 608, A8,
  \dodoi{10.1051/0004-6361/201731480}

\bibitem[{{Lujan Niemeyer} {et~al.}(2022){Lujan Niemeyer}, {Komatsu}, {Byrohl},
  {Davis}, {Fabricius}, {Gebhardt}, {Hill}, {Wisotzki}, {Bowman}, {Ciardullo},
  {Farrow}, {Finkelstein}, {Gawiser}, {Gronwall}, {Jeong}, {Landriau}, {Liu},
  {Mentuch Cooper}, {Ouchi}, {Schneider}, \&
  {Zeimann}}]{lujanniemeyer/etal2022}
{Lujan Niemeyer}, M., {Komatsu}, E., {Byrohl}, C., {et~al.} 2022, arXiv
  e-prints, arXiv:2203.04826.
\newblock \doarXiv{2203.04826}

\bibitem[{{Mas-Ribas} {et~al.}(2017){Mas-Ribas}, {Dijkstra}, {Hennawi},
  {Trenti}, {Momose}, \& {Ouchi}}]{mas-ribas/etal:2017}
{Mas-Ribas}, L., {Dijkstra}, M., {Hennawi}, J.~F., {et~al.} 2017, \apj, 841,
  19, \dodoi{10.3847/1538-4357/aa704e}

\bibitem[{{Momcheva} {et~al.}(2016){Momcheva}, {Brammer}, {van Dokkum},
  {Skelton}, {Whitaker}, {Nelson}, {Fumagalli}, {Maseda}, {Leja}, {Franx},
  {Rix}, {Bezanson}, {Da Cunha}, {Dickey}, {F{\"o}rster Schreiber},
  {Illingworth}, {Kriek}, {Labb{\'e}}, {Ulf Lange}, {Lundgren}, {Magee},
  {Marchesini}, {Oesch}, {Pacifici}, {Patel}, {Price}, {Tal}, {Wake}, {van der
  Wel}, \& {Wuyts}}]{momcheva2016}
{Momcheva}, I.~G., {Brammer}, G.~B., {van Dokkum}, P.~G., {et~al.} 2016, \apjs,
  225, 27, \dodoi{10.3847/0067-0049/225/2/27}

\bibitem[{{Nakajima} {et~al.}(2018){Nakajima}, {Fletcher}, {Ellis},
  {Robertson}, \& {Iwata}}]{nakajima/etal:2018b}
{Nakajima}, K., {Fletcher}, T., {Ellis}, R.~S., {Robertson}, B.~E., \& {Iwata},
  I. 2018, \mnras, 477, 2098, \dodoi{10.1093/mnras/sty750}

\bibitem[{{Nakajima} {et~al.}(2012){Nakajima}, {Ouchi}, {Shimasaku}, {Ono},
  {Lee}, {Foucaud}, {Ly}, {Dale}, {Salim}, {Finn}, {Almaini}, \&
  {Okamura}}]{nakajima/etal:2012}
{Nakajima}, K., {Ouchi}, M., {Shimasaku}, K., {et~al.} 2012, \apj, 745, 12,
  \dodoi{10.1088/0004-637X/745/1/12}

\bibitem[{{Ouchi}(2019)}]{ouchi:2019}
{Ouchi}, M. 2019, Saas-Fee Advanced Course, 46, 189,
  \dodoi{10.1007/978-3-662-59623-4\_3}

\bibitem[{{Planck Collaboration} {et~al.}(2020){Planck Collaboration},
  {Aghanim}, {Akrami}, {Ashdown}, {Aumont}, {Baccigalupi}, {Ballardini},
  {Banday}, {Barreiro}, {Bartolo}, {Basak}, {Battye}, {Benabed}, {Bernard},
  {Bersanelli}, {Bielewicz}, {Bock}, {Bond}, {Borrill}, {Bouchet}, {Boulanger},
  {Bucher}, {Burigana}, {Butler}, {Calabrese}, {Cardoso}, {Carron},
  {Challinor}, {Chiang}, {Chluba}, {Colombo}, {Combet}, {Contreras}, {Crill},
  {Cuttaia}, {de Bernardis}, {de Zotti}, {Delabrouille}, {Delouis}, {Di
  Valentino}, {Diego}, {Dor{\'e}}, {Douspis}, {Ducout}, {Dupac}, {Dusini},
  {Efstathiou}, {Elsner}, {En{\ss}lin}, {Eriksen}, {Fantaye}, {Farhang},
  {Fergusson}, {Fernandez-Cobos}, {Finelli}, {Forastieri}, {Frailis},
  {Fraisse}, {Franceschi}, {Frolov}, {Galeotta}, {Galli}, {Ganga},
  {G{\'e}nova-Santos}, {Gerbino}, {Ghosh}, {Gonz{\'a}lez-Nuevo}, {G{\'o}rski},
  {Gratton}, {Gruppuso}, {Gudmundsson}, {Hamann}, {Handley}, {Hansen},
  {Herranz}, {Hildebrandt}, {Hivon}, {Huang}, {Jaffe}, {Jones}, {Karakci},
  {Keih{\"a}nen}, {Keskitalo}, {Kiiveri}, {Kim}, {Kisner}, {Knox},
  {Krachmalnicoff}, {Kunz}, {Kurki-Suonio}, {Lagache}, {Lamarre}, {Lasenby},
  {Lattanzi}, {Lawrence}, {Le Jeune}, {Lemos}, {Lesgourgues}, {Levrier},
  {Lewis}, {Liguori}, {Lilje}, {Lilley}, {Lindholm}, {L{\'o}pez-Caniego},
  {Lubin}, {Ma}, {Mac{\'\i}as-P{\'e}rez}, {Maggio}, {Maino}, {Mandolesi},
  {Mangilli}, {Marcos-Caballero}, {Maris}, {Martin}, {Martinelli},
  {Mart{\'\i}nez-Gonz{\'a}lez}, {Matarrese}, {Mauri}, {McEwen}, {Meinhold},
  {Melchiorri}, {Mennella}, {Migliaccio}, {Millea}, {Mitra},
  {Miville-Desch{\^e}nes}, {Molinari}, {Montier}, {Morgante}, {Moss}, {Natoli},
  {N{\o}rgaard-Nielsen}, {Pagano}, {Paoletti}, {Partridge}, {Patanchon},
  {Peiris}, {Perrotta}, {Pettorino}, {Piacentini}, {Polastri}, {Polenta},
  {Puget}, {Rachen}, {Reinecke}, {Remazeilles}, {Renzi}, {Rocha}, {Rosset},
  {Roudier}, {Rubi{\~n}o-Mart{\'\i}n}, {Ruiz-Granados}, {Salvati}, {Sandri},
  {Savelainen}, {Scott}, {Shellard}, {Sirignano}, {Sirri}, {Spencer},
  {Sunyaev}, {Suur-Uski}, {Tauber}, {Tavagnacco}, {Tenti}, {Toffolatti},
  {Tomasi}, {Trombetti}, {Valenziano}, {Valiviita}, {Van Tent}, {Vibert},
  {Vielva}, {Villa}, {Vittorio}, {Wandelt}, {Wehus}, {White}, {White},
  {Zacchei}, \& {Zonca}}]{planck/etal:2018}
{Planck Collaboration}, {Aghanim}, N., {Akrami}, Y., {et~al.} 2020, \aap, 641,
  A6, \dodoi{10.1051/0004-6361/201833910}

\bibitem[{{Reddy} {et~al.}(2022){Reddy}, {Topping}, {Shapley}, {Steidel},
  {Sanders}, {Du}, {Coil}, {Mobasher}, {Price}, \& {Shivaei}}]{reddy/etal:2022}
{Reddy}, N.~A., {Topping}, M.~W., {Shapley}, A.~E., {et~al.} 2022, \apj, 926,
  31, \dodoi{10.3847/1538-4357/ac3b4c}

\bibitem[{{Rowan-Robinson} {et~al.}(2016){Rowan-Robinson}, {Oliver}, {Wang},
  {Farrah}, {Clements}, {Gruppioni}, {Marchetti}, {Rigopoulou}, \&
  {Vaccari}}]{rowan-robinson/etal:2016}
{Rowan-Robinson}, M., {Oliver}, S., {Wang}, L., {et~al.} 2016, \mnras, 461,
  1100, \dodoi{10.1093/mnras/stw1169}

\bibitem[{{Runnholm} {et~al.}(2020){Runnholm}, {Hayes}, {Melinder},
  {Rivera-Thorsen}, {{\"O}stlin}, {Cannon}, \& {Kunth}}]{runnholm/etal:2020}
{Runnholm}, A., {Hayes}, M., {Melinder}, J., {et~al.} 2020, \apj, 892, 48,
  \dodoi{10.3847/1538-4357/ab7a91}

\bibitem[{{Shimakawa} {et~al.}(2017){Shimakawa}, {Kodama}, {Shibuya},
  {Kashikawa}, {Tanaka}, {Matsuda}, {Tadaki}, {Koyama}, {Hayashi}, {Suzuki}, \&
  {Yamamoto}}]{shimakawa2017}
{Shimakawa}, R., {Kodama}, T., {Shibuya}, T., {et~al.} 2017, \mnras, 468, 1123,
  \dodoi{10.1093/mnras/stx091}

\bibitem[{{Steidel} {et~al.}(2011){Steidel}, {Bogosavljevi{\'c}}, {Shapley},
  {Kollmeier}, {Reddy}, {Erb}, \& {Pettini}}]{steidel/etal:2011}
{Steidel}, C.~C., {Bogosavljevi{\'c}}, M., {Shapley}, A.~E., {et~al.} 2011,
  \apj, 736, 160, \dodoi{10.1088/0004-637X/736/2/160}

\bibitem[{{Trainor} {et~al.}(2016){Trainor}, {Strom}, {Steidel}, \&
  {Rudie}}]{trainor/etal:2016}
{Trainor}, R.~F., {Strom}, A.~L., {Steidel}, C.~C., \& {Rudie}, G.~C. 2016,
  \apj, 832, 171, \dodoi{10.3847/0004-637X/832/2/171}

\bibitem[{{Trainor} {et~al.}(2019){Trainor}, {Strom}, {Steidel}, {Rudie},
  {Chen}, \& {Theios}}]{trainor/etal:2019}
{Trainor}, R.~F., {Strom}, A.~L., {Steidel}, C.~C., {et~al.} 2019, \apj, 887,
  85, \dodoi{10.3847/1538-4357/ab4993}

\bibitem[{{Verhamme} {et~al.}(2006){Verhamme}, {Schaerer}, \&
  {Maselli}}]{verhamme/etal:2006}
{Verhamme}, A., {Schaerer}, D., \& {Maselli}, A. 2006, \aap, 460, 397,
  \dodoi{10.1051/0004-6361:20065554}

\bibitem[{Virtanen {et~al.}(2020)Virtanen, Gommers, Oliphant, Haberland, Reddy,
  Cournapeau, Burovski, Peterson, Weckesser, Bright, {van der Walt}, Brett,
  Wilson, Millman, Mayorov, Nelson, Jones, Kern, Larson, Carey, Polat, Feng,
  Moore, {VanderPlas}, Laxalde, Perktold, Cimrman, Henriksen, Quintero, Harris,
  Archibald, Ribeiro, Pedregosa, {van Mulbregt}, \& {SciPy 1.0
  Contributors}}]{2020SciPy-NMeth}
Virtanen, P., Gommers, R., Oliphant, T.~E., {et~al.} 2020, Nature Methods, 17,
  261, \dodoi{10.1038/s41592-019-0686-2}

\bibitem[{{Weiss} {et~al.}(2021){Weiss}, {Bowman}, {Ciardullo}, {Zeimann},
  {Gronwall}, {Mentuch Cooper}, {Gebhardt}, {Hill}, {Blanc}, {Farrow},
  {Finkelstein}, {Gawiser}, {Janowiecki}, {Jogee}, {Schneider}, \&
  {Wisotzki}}]{weiss2021}
{Weiss}, L.~H., {Bowman}, W.~P., {Ciardullo}, R., {et~al.} 2021, \apj, 912,
  100, \dodoi{10.3847/1538-4357/abedb9}

\bibitem[{{Wisotzki} {et~al.}(2016){Wisotzki}, {Bacon}, {Blaizot},
  {Brinchmann}, {Herenz}, {Schaye}, {Bouch{\'e}}, {Cantalupo}, {Contini},
  {Carollo}, {Caruana}, {Courbot}, {Emsellem}, {Kamann}, {Kerutt}, {Leclercq},
  {Lilly}, {Patr{\'\i}cio}, {Sandin}, {Steinmetz}, {Straka}, {Urrutia},
  {Verhamme}, {Weilbacher}, \& {Wendt}}]{wisotzki/etal:2016}
{Wisotzki}, L., {Bacon}, R., {Blaizot}, J., {et~al.} 2016, \aap, 587, A98,
  \dodoi{10.1051/0004-6361/201527384}

\bibitem[{{Wisotzki} {et~al.}(2018){Wisotzki}, {Bacon}, {Brinchmann},
  {Cantalupo}, {Richter}, {Schaye}, {Schmidt}, {Urrutia}, {Weilbacher},
  {Akhlaghi}, {Bouch{\'e}}, {Contini}, {Guiderdoni}, {Herenz}, {Inami},
  {Kerutt}, {Leclercq}, {Marino}, {Maseda}, {Monreal-Ibero}, {Nanayakkara},
  {Richard}, {Saust}, {Steinmetz}, \& {Wendt}}]{wisotzki/etal:2018}
{Wisotzki}, L., {Bacon}, R., {Brinchmann}, J., {et~al.} 2018, \nat, 562, 229,
  \dodoi{10.1038/s41586-018-0564-6}

\end{thebibliography}
\bibliographystyle{aasjournal}

\end{document}